\documentstyle[12pt,aasms4,psfig]{article}
\slugcomment{Accepted in AJ, January 09, 2002}
\lefthead{Garc\'ia-Barreto, Franco \& Rudnick}
\righthead{VLA High Resolution 1.4 and 8.4 GHz Mapping of the Barred
Galaxy NGC 3367}

\begin{document}

\title{VLA High Resolution 1.4 and 8.4 GHz Mapping of the Barred
Galaxy NGC 3367}

\author{J. Antonio Garc\'{\i}a-Barreto, Jos\'e Franco}
\affil{Instituto de Astronom\'{\i}a,
Universidad Nacional Aut\'onoma de M\'exico, Apartado Postal 70-264,
M\'exico D.F. 04510 M\'exico}

\and

\author{Lawrence Rudnick}
\affil{Department of Astronomy, University of Minnesota, 116 Church St.,
S.E., Minneapolis, MN 55455}

\begin{abstract}
We report new radio continuum observations with an angular resolution
of 2$''.1$ at 1.4 GHz (20 cm) and $0''.28$ at 8.4 GHz (3.6 cm), of the
barred galaxy NGC 3367. In the map at 1.4 GHz, the central nuclear
region connects to the SW lobe, with a projected structure of at a position angle of
P.A.$\sim230^{\circ}$, forming a jet-like structure.  The map at 8.4
GHz shows a compact unresolved source (smaller than $65\,$ pc in diameter)
associated with emission from the nucleus and several compact sources
located within a radius of about 300 pc, forming a circumnuclear structure.
The compact core, jet, and lobes form a small, low-power counterpart to
radio galaxies, with a flow axis that is out of the plane of the galaxy.
The flow axis (P.A.$\sim230^{\circ}$) coincides with the P.A. of
the major axis of the galaxy and is thus inclined to the rotation axis of the disk.
In addition, the flow axis differs by about 20$^{\circ}$ from the major axis of
the stellar bar. Assuming that the stellar bar rotates counterclockwise
(ie. assuming trailing spiral arms), this difference in angle is taken as an argument
in favor of having the jet-like structure out of the plane of the disk
and not associated with the stellar bar.

\end{abstract}

\section{Introduction}

Strong radio continuum emitters, such as radio galaxies or quasars,
are identified with elliptical galaxies and recent mergers, and many
of them have impressive radio jets with sizes larger than the host
galaxy (see Wilson \& Colbert 1995).  Maps of radio galaxies display
fairly straight radio jets with distant lobes, showing the interaction
of the jet with the ambient medium and indicating that the jet axis
has been stable for thousands or even millions of years. Recent
work indicates the presence of lobes at a distance of hundred kiloparsecs
from a spiral galaxy in a cluster (\cite{led98,led01}). In normal spiral
galaxies, on the other hand, most of the radio continuum emission comes from the
disk component (\cite{hum81,con87,gar93,nik97}). Radio surveys of Seyfert galaxies indicate
that the radio continuum emission arises mainly from three components:
(1) subkiloparsec emission from the nuclear region; (2) extranuclear
kpc scale emission; and (3) greater than kpc-scale emission associated
with the disk (\cite{wil83,ulv84,bau93,col96,ho01}). In barred spiral galaxies,
the radio continuum emission is found from the following: (1) emission from the
compact nucleus; (2) emission from circumnuclear region ($\leq 1$kpc);
(3) emission from dust lanes in leading side of the stellar bar; and (4) emission
from spirals arms and disk (\cite{hum87,con87,gar91a,gar91b,lin99,bec99}).
Depending on the central activity, a spiral can be
classified as a starburst or an active galaxy. Seyfert galaxies are stronger
radio emitters at 1.4 GHz than normal and barred spirals
(\cite{con87}).  The central radio sources are
sometimes associated with a pair of extended sources identified as
lobes (\cite{ulv84,ulv87,bau93,col96,ho01}).  These triple sources (compact
nucleus plus lobes) are thought to be small scale, low power versions
of the large scale jets and lobes seen in radio galaxies and quasars. The radio
continuum emission from the central region of spirals is either linked to star
formation, via HII regions and supernova remnants as in starbursts, or to an
unresolved compact object, probably an AGN (Baum et al 1993). The study of
NGC 3367 may be important in order to understand the lobe-central source relationship
and help to clarify the starburst-AGN dichotomy.

The barred spiral, NGC 3367, is a mildly active galaxy, between a weak
Liner and a HII nucleus (\cite{ver97,ho97}) that displays such a triple source
structure. At 15$''$ angular resolution, in addition to extended
emission from different locations in the disk, it shows an unresolved
radio source in the center plus two sources in opposite directions from it
(\cite{con90}). Maps at 4$''.5$ resolution indicated more clearly the presence
of the two lobes, (\cite{gar98}). The central source was not resolved
at this 4$''.5$ angular resolution, and it was unclear if it is a
point-like source or an extended circumnuclear structure, and whether
or not it is really connected with the lobes.

In this paper we present new VLA\footnote{The VLA is part of the
National Radio Astronomy Observatory which is a facility of the
National Science Foundation operated under agreement by Associated
Universities Inc.} radio continuum observations of NGC~3367 at 1.4 GHz
with a beam of $\approx2''.1$, and at 8.4 GHz with a beam of
$\approx0''.28$. \S 2 presents the observed
properties of NGC 3367, \S 3 presents the new radio continuum
observations and results, and \S 4 gives the discussion
and conclusions.

\section {NGC 3367}

NGC~3367 is an SBc(s) barred spiral galaxy with a stellar bar structure
of diameter $\approx 32''$ (6.7 kpc) oriented at a position angle (PA)
$\approx 70^{\circ}$. The disk of NGC 3367 is inclined with respect
to the plane of the sky at an angle between $6^{\circ}$ (Grosb\o l
1985) and $30^{\circ}$ (\cite{gar01}), and has an optically bright SW
structure resembling a half-ring, or a large scale ``bow shock'', at
about 10 kpc from the nucleus.  This structure is formed by a
collection of H$\alpha$ regions, that looks like a necklace and its origin has
been ascribed to an off-center impact with an external intruder, most likely a small
galaxy (\cite{gar96a}).  Indeed, the general
arrangement of the H$\alpha$ knots is similar to the elongated rings
found in numerical simulations of off-center galaxy collisions by
Gerber \& Lamb (1994).  The expanding density wave created by the
collision can trigger the formation of the half-ring of HII regions
and, given that the expected wave velocity in the disk is below 100
km/s, the collision probably ocurred a few times 10$^8$ yr ago.  In
addition, aside from the radial gas inflows induced by the stellar
bar, a galactic collision is also able to drive gas towards the
galactic center inducing circumnuclear star formation as well as
nuclear activity. NGC 3367 also shows H$\alpha$ emission from the
central region with an unresolved source, most likely, a combination
of emission from a compact source and circumnuclear structure at a
radius of less than 500 pc (\cite{gar96a,gar96b}).

The disk of NGC 3367 has a normal content of atomic hydrogen, with
M$_{HI}\sim 7\times10^9$ M$_{\odot}$ (\cite{huc85}), and it is
considered an isolated field galaxy, behind the Leo group of galaxies at a
distance of 43.6 Mpc, with its closest neighbor more than 900 kpc away
to the NE (\cite{tul88}).  Its optical spectrum shows moderately broad
H$\alpha$+[NII] lines with FWHM$\sim650$ km s$^{-1}$, but H$\beta$ is
stronger than [OIII]$\lambda 5007$\AA~ and there is weak emission of
He II $\lambda 4686$\AA, suggesting the existence of WR stars and
weak Liner activity (\cite{ver86,ver97,ho97}). In addition, its X~ray
luminosity is stronger than that of any normal spiral galaxy, but weaker than
Seyfert or radio galaxies (\cite{gio90,sto91,fab92}).  Nonetheless, from
the optical line ratios, NGC 3367 is also considered to have an HII nucleus
(\cite{ho97}).

The first radio continuum observation of the region around NGC 3367 was at
178 MHz, with several arcmin angular resolution (\cite{gow67}). Although
the radio emission was identified with NGC 3367 (\cite{cas67}), the bulk of
the emission most likely originated from a radio galaxy $\approx3'$ north
of NGC 3367, detected later with better angular resolution (\cite{law83}).
Similarly, observations from Arecibo at 430 MHz and 835 MHz with integrated fluxes of
583 mJy and 365 mJy respectively with $\approx9'$ angular resolution
most likely included the flux of the background radio galaxy (\cite{isr83}).
Green Bank single dish observations of NGC 3367 at 5 GHz with a resolution
of $\sim3'$ were carried out by Sramek (1975) and Bennett et al. (1986) with
integrated fluxes of 35 mJy and 71 mJy respectively. Israel \& van der Hulst (1983)
reported an integrated flux of 18 mJy at 10.7 GHz with an angular resolution
of $3'$ from OVRO 40m. Finally Dunne et al. (2000) reported a flux of 132 mJy at
350 GHz with $15''$ angular resolution.

The first aperture synthesis map of NGC 3367 at 1.49 GHz, with 15$''$ angular
resolution, showed diffuse disk emission plus three peaks aligned in
the NE -- SW direction (\cite{con90}), and the disk emission was
noticed to be edge brightened.  A more recent mapping at this same
frequency, but now with 4$.''5$ angular resolution, shows more clearly
the emission from the triple source: emission from the nuclear region in addition
to emission from two extended lobes at a distance of $\sim 6$ kpc from the
center (\cite{gar98}).  The polarization analysis indicated that only
the SW lobe is polarized, suggesting that it is out of the disk of the
galaxy and closer to the observer than the NE lobe (the emission of the NE lobe
has been depolarized because this emission has passed through
the plane of the galaxy)(\cite{gar98}). These observations showed very
clearly the presence of kiloparsec scale lobes from a barred spiral
galaxy seen almost face on, in addition to the weaker emission from
a large number of compact sources in the disk.

\section{Observations \& Results}

We have carried out radio continuum observations at the VLA in New
Mexico in the A array at 1.3851 GHz, 1.4649 GHz, 8.4351 GHz and 8.4851
GHz in 1998 April 23, using 27 antennas with 50 MHz bandwidths, and
$\approx3^h$ integration time at 1.4 GHz and $\approx4^h$ at 8.4 GHz
on NGC 3367.  We observed 3C286 as the amplitude and polarization
calibrator and 1120+143 as the phase calibrator at 1.4 GHz and
1051+213 as the phase calibrator at 8.4 GHz.  We adopted a flux
density of 3C286 of S$_{\nu}=14.554$ Jy at ${\nu}=1.4649$ GHz, and
S$_{\nu}= 5.1702$ Jy at ${\nu}=8.4851$ GHz.  Several iterations of
phase self-calibration were used at 1.4 GHz.  No self-calibration was
used at 8.4 GHz because the peak emission was only $\approx1$ mJy
beam$^{-1}$.

\begin{figure}[th]
\psfig{figure=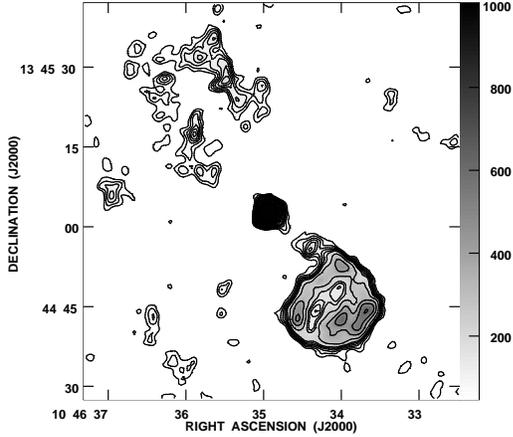,width=7cm,angle=0}
\caption[figure=garcia3367.fig1.ps]{Radio continuum emission from NGC 3367 at 1.4 GHz with a
beam FWHM$\approx 2''.1\times1''.8$ at PA$\approx-67^{\circ}$. The
contours are in units of 25$\mu$Jy/beam and the levels are
2, 3, 4, 5, 6, 7, 10, 15, 20, 25, 30, 50, 65, 80, and 120. The weak sources
outside the area of the central region plus lobes are real and belong
to emission from the disk, as can be seen in lower resolution maps
shown in Figures 2 and 6 of Garcia-Barreto et al. (1998). Greyscale
is from 50 $\mu$Jy/beam to 1 mJy/beam}
\label{fig1}
\label{hi}
\end{figure}

\subsection{The Jets and the SW Lobe}

Figure 1 shows the total intensity map at 1.4 GHz of NGC 3367 with a
full width at half maximum (FWHM) angular resolution beam of
$2.''1\times1''.8$ at a P.A.$\approx-67^{\circ}$ and an rms noise of
$\sim25~{\mu}$Jy beam$^{-1}$.  The emission above
the 2$\sigma$ level clearly shows the central source and lobes.  The
central source has a peak emission of 11.2 mJy beam$^{-1}$ and is
still unresolved.  At the 2$\sigma$ level, there is no clear connection of
the central source and the SW lobe.  There are, however, extensions
pointing toward the NE and SW lobes, and a faint connection appears at
about 1$\sigma$ (see below).  At this new higher angular resolution
1.4 GHz observation, the emission from the lobes starts to be
resolved, and the SW lobe shows the same polarization (not shown here)
as in the previous $4''.5$ angular resolution observations, namely that
the eastern side of the lobe shows the strongest polarization
(\cite{gar98}).

\begin{figure}[th]
\psfig{figure=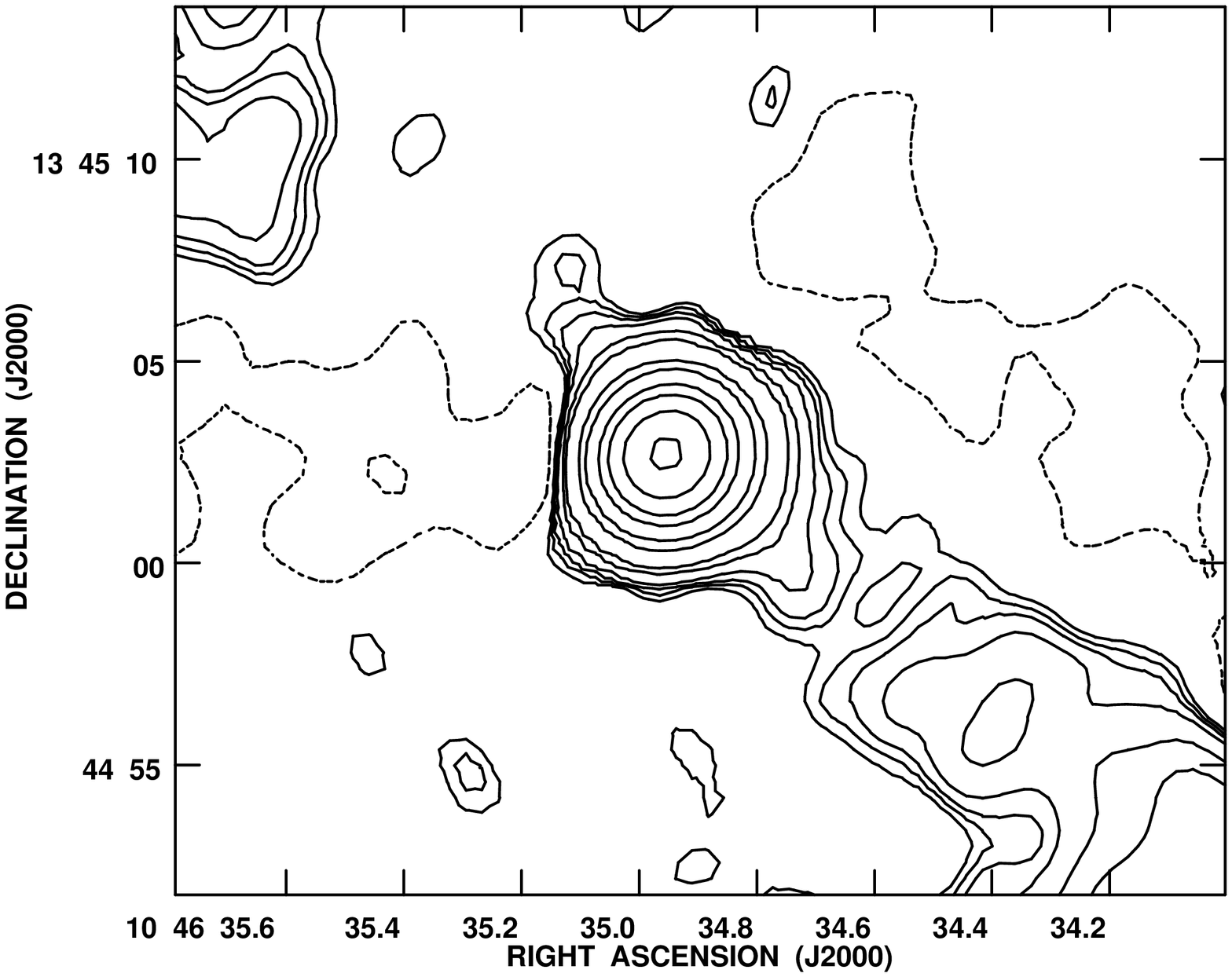,width=8cm,angle=-90}
\caption[figure=garcia3367.fig2.ps]{Radio continuum emission from the innermost
central region at 1.4 GHz from Figure 1. The position angle (P.A.) of the
radio continuum emission connecting the central source with the south west
(SW) lobe is P.A$._{rad}\sim230^{\circ}\pm10^{\circ}$; the stellar bar P.A. is
P.A$._{bar}\sim245^{\circ}\pm5^{\circ}$; the approaching semi major kinematical
axis is P.A$._{ma}=231^{\circ}$. The contours are in units of
25$\mu$Jy/beam and the levels are -2, 1, 1.5, 2, 3, 6, 10, 30, 50, 100, 150, 250 and
410. The second negative contour left of the central emission only indicates
a relative maximum ($\approx -40 \mu$Jy/beam) and not a deeper region.}
\label{fig2}
\label{hi}
\end{figure}

The emission from both lobes is mostly diffuse, with some clumpy
structure.  The morphology of the SW lobe resembles a bow shock,
probably due to the interaction of a relativistic jet with a low
density ambient medium.  There is faint emission directed from the central
radio source towards the SW lobe, indicating that they are
connected through a low surface brightness continuous structure, $7''\times4''$
(2.5 kpc $\times$ 850 pc), perhaps jet-like. The width is probably an upper limit
dictated by the restoring beam size. This is
seen in Figure 2, which shows the faint emission from the innermost $25''$
central region of NGC 3367 at 1.4 GHz.  The contours are drawn starting at a 1$\sigma$
level, and shows that the central emission indeed connects to the SW
lobe. Although in Figure 2 there is a minimum of emission midway
(ie. $\alpha \approx34^s.6; \delta \approx44'~59''$) the 1$\sigma$ contour surrounds this
minimum and connects the central radio emission with the SW lobe; in order to verify
the existence of the structure, we did make several maps with different restoring beams
with self calibration of phase and amplitude and the structure was there
in all maps although with slightly different detailed morphology (not shown here).
Therefore we feel confident that the structure connecting the central radio emission
and the SW lobe really exists; however, the detailed morphology needs to be taken with
caution. The average P.A.$_{rad}$ of this structure is $\sim230^{\circ}$ with a spread
of $\pm10^{\circ}$.  The NE extension of the central radio source is seen clearly at a P.A.
$\sim35^{\circ}$, and seems to be the starting part of the counter jet
that flows towards the NE lobe. The stellar bar, however, is oriented at
P.A.$_{bar}\sim245^{\circ}\pm5^{\circ}$. In the SW, the radio extension lies to the
south of the optical bar, while in the NE, the radio extension lies to the north
of the bar (see Figure 3). There is no radio continuum emission (at the three sigma level of
the rms noise) from any dust lane in the leading side of the stellar bar, as
is often detected in other barred galaxies (\cite{ond83,bec99}). The fact that the
SW lobe is out of the plane of the galaxy suggests that the relative alignment
between the radio structure (connecting the central radio source and the SW lobe)
and the stellar bar may be only a projection of the structure onto the galaxy disk
and that the structure is out of the plane of the galaxy disk.

\begin{figure}[t]
\psfig{figure=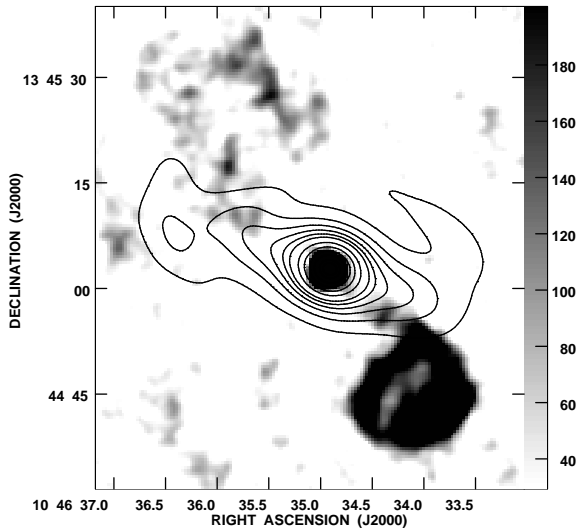,width=8cm,angle=0}
\caption[figure=garcia3367.fig3.ps]{Radio continuum emission at 1.4 GHz, in greyscale and the
optical continuum (I broadband filter centered
at $\lambda$8040 \AA~) in contours. The contours are in arbitrary
units relative to peak showing mainly the bright
emission from the central stellar bar.}
\label{fig3}
\label{hi}
\end{figure}

\subsection{Compact Nucleus and Circumnuclear Structure}

Figure 4 shows the radio continuum contours of the emission from the
innermost $4''$ central region of the galaxy at 8.4 GHz, with an
angular resolution of $0''.28\times0''.25$ at a P.A.$\sim1^{\circ}$ and
an rms noise of 11$\mu$Jy beam$^{-1}$.  The emission is dominated
by a still unresolved central source of 0.96 mJy beam$^{-1}$, at
$\alpha(J2000)=10^h46^m34^s.956~~\delta(J2000)
=+13^{\circ}45'02''.94$.  Its deconvolved diameter is less than 65 pc,
and is surrounded by several low surface brightness peaks of emission
out to a distance of $\approx300$ pc.  A rather similar structure is
found in the lower resolution map at 1.4 GHz when a central unresolved
source is subtracted.  At 8.4 GHz, there is a short extension at a PA
of 230$^{\circ}$, the same as seen for the jet on larger scales.  No
polarization is detected at 8.4 GHz, with an upper limit of 2 \% for
the compact nucleus, and a characteristic limit of 25 \% for the
circumnuclear structure. Peak fluxes at various positions in the
circumnuclear region are all smaller than 100$\mu$Jy beam$^{-1}$.

\begin{figure}[t]
\psfig{figure=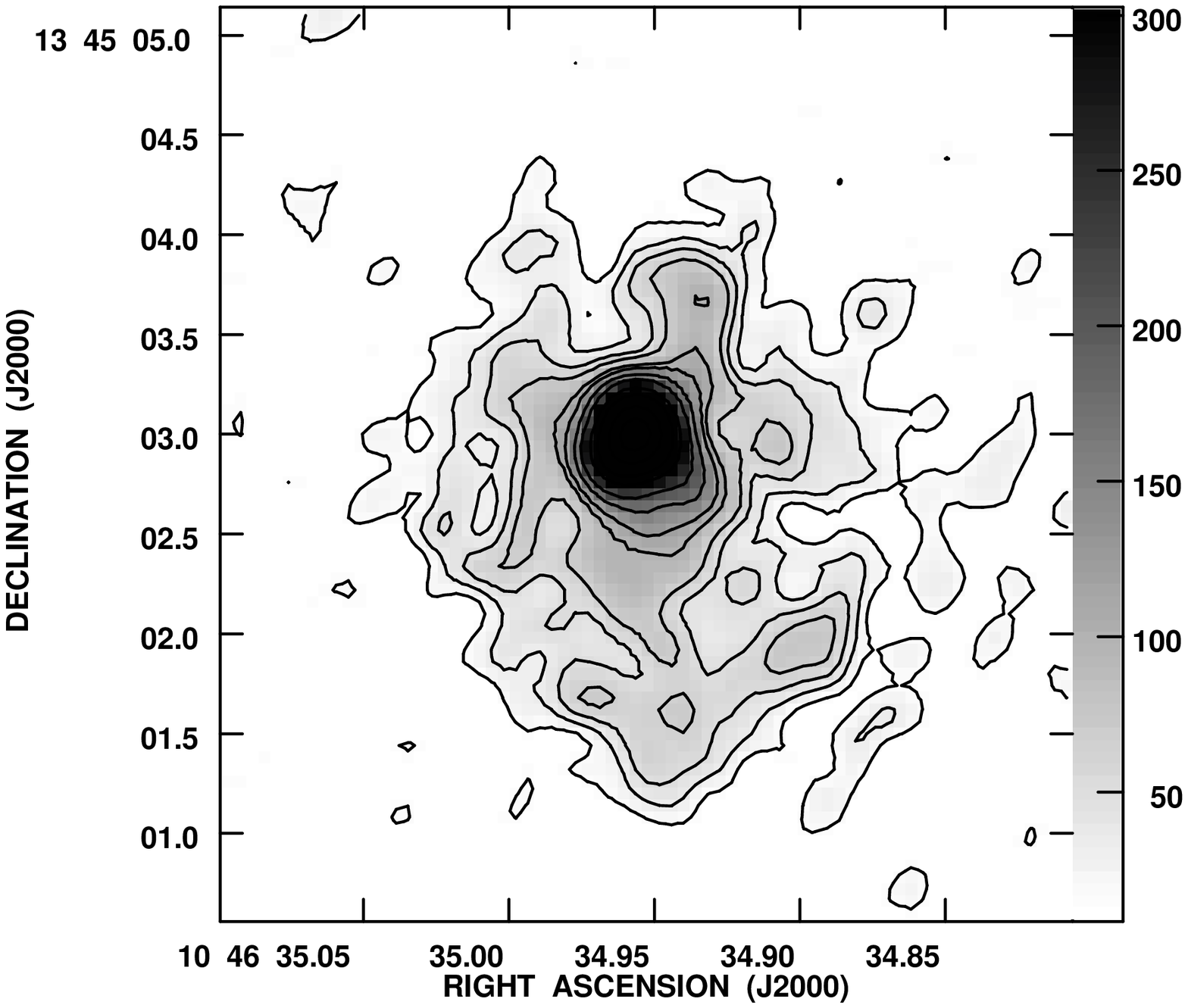,width=8cm,angle=0}
\caption[figure=garcia3367.fig4.ps]{Radio continuum emission from the innermost central
region at 8.4 GHz. The restoring beam FWHM is $0''.28\times0''.25$ ($\sim60$ pc).
The contours are in units of 11$\mu$Jy/beam and the levels are -3, 1.5, 3, 4.5, 6, 9, 11,
15, 20, 30, 40, 60, 80 and 100. Greyscale is from 11 to 300 $\mu$Jy beam$^{-1}$.}
\label{fig4}
\label{hi}
\end{figure}

        Figure 5 shows the innermost $10''$ central radio continuum with a
superposition of the 1.4 GHz (in contours) and the 8.4 GHz (in
greyscale) emissions.  The emission is clearly dominated by the
central source and an estimate of the spectral index between 1.4 GHz
and 8.4 GHz (integrated over the central $3.5''$) gives
$\alpha\sim-0.53$ (S$_{\nu}\propto\nu^{\alpha}$), indicating that the
radiation is mainly synchrotron emission.  The interpretation of
spectral index should be taken with caution since the fluxes are the
sum of the core plus the circumnuclear sources. The total flux at 1.4 GHz
within $7''$ is 16.8 mJy, while the total flux at 8.4 GHz within $4''.5$
is 5.4 mJy.

\section{Discussion}

We have observed the radio continuum emission from the barred galaxy
NGC~3367, with the VLA A array, at 1.4 GHz and 8.4 GHz with 2$''.1$
and $0''.28$ spatial resolutions, respectively.  The radio maps show
emission from the central region, the lobes, and a weak jet.
This morphology in NGC 3367 resembles the morphology of more powerful radio
galaxies (\cite{far74,sch01}). The power at 1.4 GHz is two to three orders of
magnitude less than any radio galaxy and it is more like the power of Seyfert
galaxies, see Table 1 (see Figure 3 of \cite{ho01,ulv01}). Nonetheless this is one of the
largest and best defined triple source ever detected in a galaxy considered to
be a normal barred spiral. The triple source is due to emission from the central sources,
emission from a jet-like structure connecting the central emission with
the lobes and the extended lobes. Other large radio structures have been observed
in the Seyfert galaxies , for example, Mrk 6 (lobe extent 14 kpc), Mrk 348 (lobe extent 5 kpc),
NGC 3516 (lobe extent 8.5 kpc) (\cite{bau93}), NGC 4235 (lobe extent 9 kpc) (\cite{col96})
and in the disk galaxy O313-192 in the cluster A428 (lobe extent 100 kpc) (\cite{led98,led01}).
The radio continuum jets in other barred galaxies as NGC 1068 (\cite{wil87}) and NGC 5728 (\cite{sch88})
are much smaller in size (of the order of tens to hundred parsecs). In contrast
with O313-192, which has an AGN, and the Markarian galaxies, which are Seyfert 1s and 2s,
NGC 3367 is only a mildly active Liner and, based on optical line emission ratios and widths, is
not even considered to be a Seyfert galaxy.

\begin{figure}[t]
\psfig{figure=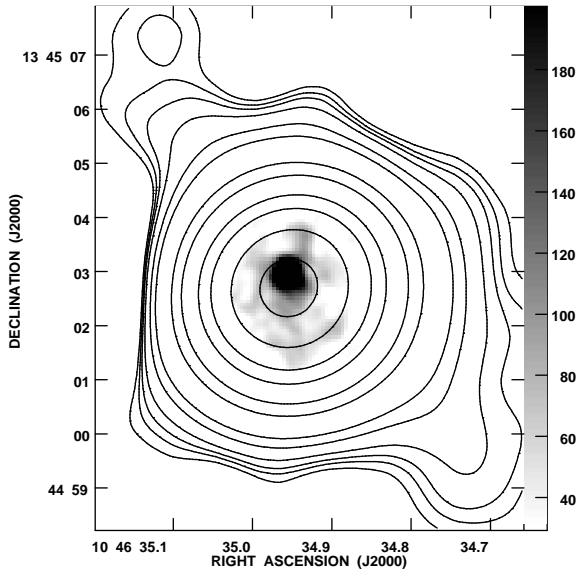,width=8cm,angle=0}
\caption[figure=garcia3367.fig5.ps]{Radio continuum emission from the innermost central
region at 1.4 GHz, in contours, and at 8.4 GHz, in greyscale. The contours are
in units of 25$\mu$Jy/beam and the levels are 1, 1.5, 2, 3, 6, 10, 30, 50, 100, 150,
250 and 390. The greyscale is from 30$\mu$Jy/beam to 200$\mu$Jy/beam.}
\label{fig5}
\label{hi}
\end{figure}

The P.A$._{jet}$ of the jet-like structure ($\sim230^{\circ}\pm10^{\circ}$; see Figure 3)
is the same as the P.A$._{ma}$ of the major axis, as determined from H$\alpha$
kinematics (\cite{gar01}).  The jet would be considered a low power one (\cite{mas96}).
The direction of the plasma outflow is thus not aligned with the rotation axis
of the disk because, if that were the case, the relative orientation of the
lobes, in NGC 3367, would have to lie closer to the direction SE-NW, that is, the
projection of the P.A. of the minor axis as expected in the very simple picture
that jets emanating from an active nucleus would emerge at right angles to the
disk of the host galaxies (\cite{kin00}).  However the simple scenario is
contradicted by the observations of Seyfert galaxies (\cite{sch97,kin00,ulv01}).
Our observations of NGC 3367 suggest that the outflow axis is inclined with respect to
the rotation axis of the galaxy and also inclined to the line of sight. A comparison
between the the P.A. of the extended radio structures from
Seyferts with their host galaxies' major-axis P.A. indicate that
the radio structures in type-2 Seyferts are oriented along any direction
in the galaxy, and not necessarily along the minor axis (\cite{sch97,kin00,ulv01}).
The directions of the radio jets are consistent with being completely
uncorrelated with the planes of the host galaxies (\cite{pri99,nag99,kin00}).
Our observations indicate that NGC 3367 (being a noninteracting late-type
Liner/HII spiral having kpc lobes) presents P.A$_{radio}\approx$P.A$_{maj}$.

\begin{figure}[t]
\psfig{figure=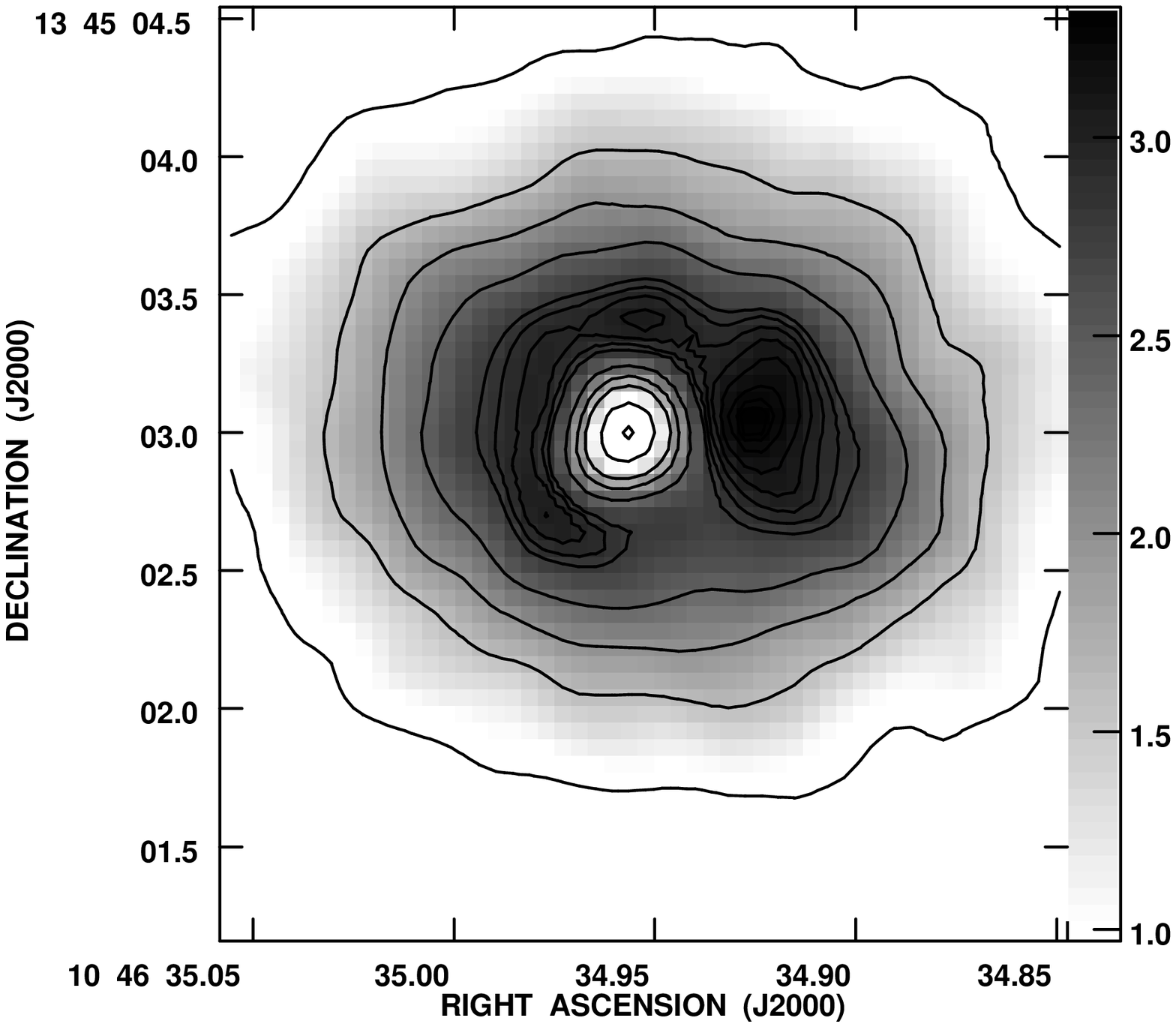,width=8cm,angle=0}
\caption[figure=garcia3367.fig6.ps]{H$\alpha$ - 8.4 GHz emission from the innermost $4''$. The contours and
greyscale are in arbitrary units relative to the maximum. The structure observed
indeed suggests the existence of a circumnuclear
structure within the innermost 300 pc ($1''=210$ pc). This image was obtained with
several assumptions, among them are (1) the spatial location of the peak of the
H$\alpha$ emission coincides with the spatial location of the 8.4 GHz emission;
(2) both, the peaks of H$\alpha$ and 8.4 GHz emissions are directly proportional
to each other; (3) the constant of proportionality was chosen as to have zero emission
from the center. Although the assumptions seem reasonable, the structure is definitely
not a proof, but it is very suggestive, of the existence of a circumnuclear structure
in H$\alpha$.}
\label{fig6}
\label{hi}
\end{figure}

        The emission at 8.4 GHz, with the higher resolution, is complex and
shows an unresolved nuclear source and several sources surrounding it.
The circumnuclear radio sources are at all position angles (see Figs.
4 and 5), and do not align at all with the lobes or with any other
optical structure in the galaxy.  This implies that the circumnuclear sources
are probably located in the plane of the disk, and might be identified with a
circumnuclear structure at distances of 125 pc to 325 pc.  The radio emission
from the inner $4''.5$ is then most likely a mixture of synchrotron and free-free
emission and the sources are most probably giant HII regions near an
inner Lindblad resonance (as it is the case in other barred galaxies).

        An additional hint for the existence of a circumnuclear structure is provided
by the H$\alpha$ emission. As mentioned earlier NGC 3367 shows H$\alpha$ emission
from the central region which is unresolved from ground observations
(\cite{gar96a,gar96b}) with a bright emission from a compact source and weak
extended emission from within $6''$. Assuming that the peak of H$\alpha$ coincides
with the peak of the 8.4 GHz radio continuum emission, we then have subtracted
the 8.4 GHz radio continuum emission from the H$\alpha$ emission in such a way
as to have zero emission from the center. The image, H$\alpha$ - 8.4 GHz radio continuum,
that is produced is shown in Figure 6; it shows an H$\alpha$ circumnuclear structure.
This H$\alpha$ image provides {\it no} proof of the existence of a circumnuclear structure
since we have made several assumptions: (1) that the spatial position of peak of
the H$\alpha$ emission coincides with the position of the 8.4 GHz emission; (2) that
the radio continuum 8.4 GHz is related to the H$\alpha$ emission in such a way that they
are directly proportional to each other through a constant; (3) that the constant of
proportionality was chosen in such a way as to have zero emission from the very center;
(4) that the final subtracted image represents emission from the disk of the galaxy
where we think the structure lies (near a ILR). The image however is very suggestive
of the existence of such a structure. If the circumnuclear structure represents
regions of massive star formation one could estimate the rate of supernova using
the relation  (\cite{con92}):

\begin{equation}
\frac {\nu_{SN}}{yr^{-1}}  \sim \frac {L_N / (10^{22} W Hz^{-1}) }  {13 (\nu /GHz)^{-\alpha} }
\end{equation}

using the integrated flux within the innermost $3''.5$ at 1.4 GHz and
assuming that all of this emission is of non-thermal origin (which we know is wrong
but hope that not by much) and a spectral index $\alpha=-0.53$ then the expected
supernova rate in the circumnuclear region of NGC 3367 is $\nu_{SN}=0.03$, which
is very similar for other galaxies (\cite{con92}). The global star formation rate
using the far infrared ({\it IRAS}) luminosity and using the relation (\cite{con92}):

\begin{equation}
\frac {SFR(M\geq5 M_{\odot})}{M_{\odot}yr^{-1}} \sim \frac{L_{FIR}/L_{\odot}}{1.1\times10^{10}}
\end{equation}

with a L$_{FIR}=2\times10^{10} L_{\odot}$ (\cite{soi89}) we get
SFR$_{NGC 3367}\sim1.8 M_{\odot}~yr^{-1}$ a value which similar to the values found in
other galaxies (\cite{con92}).

        As stated above, the gas in the circumnuclear structure could have been
driven inwards either by the perturbations induced by the potential of the
stellar bar, or by a possible off-center collision with a minor
intruder. A reason in favor of the active compact radio nucleus is
the short extension in the 8.4 GHz map that has a P.A. similar to that of the
jet-like structure observed in the 1.4 GHz map that connects the central
radio source with the SW lobe. If the flow were coming from a starburst
wind (originating from the circumnuclear structure), this wind would be directed
(in the most simple case) out of the plane towards the rotation axis of the disk,
(where the density gradient is largest) and the jet - like
and lobes would be projected in the SE - NW direction as mentioned above.
Therefore, although an elongated structure is not observed at 8.4 GHz forming
a jet, the observations suggest that the plasma flows out from the compact
nucleus, possibly directed in the observed orientation as a result of an
accretion disk inclined (but not perpendicular) with respect to the plane of
the galaxy. One thus could infer that the jet does not interact with the circumnuclear material.
The interpretation of the origin of the energy for the jets, in NGC 3367, might still
controversial in the following respect, namely, the global q parameter. The
q parameter is the ratio of the FIR ({\it IRAS}) emission to the 1.4 GHz
emission (\cite{hel85}). Empirically the median $2.2 \leq \langle q \rangle \leq 3.1$ was found
for spiral galaxies (\cite{con95}), while $\langle q \rangle \leq 2$ was found for
galaxies powered by an AGN (\cite{con91b}). In particular, based on the
value found for q for NGC 3367 (q$=2.4$ using the 5 GHz total emission)
it was concluded that the dominant energy source in NGC 3367 is stars (\cite{con91b,con95}).
If we compute q using the total emission at 1.4 GHz (119 mJy [\cite{con98}]),
we get is q=1.9 which might indicate the presence of an (mildly) AGN according to the
convention (\cite{con91b,con95}). We believe that this result involves global emissions
(radio and infrared) and not necessarily the energy powering the jets and lobes observed
in NGC 3367 (\cite{con91b,con95}). This result still needs to be confronted
with the current observations of the central radio sources, jet and lobes in NGC 3367.

This central radio continuum structure (compact source with
circumnuclear regions) is very similar to the structure observed in
some barred Sy 1 galaxies, like NGC 1097 (\cite{hum87}) and NGC 7469
(\cite{con91a,wil91b,mil94,mau94,gen95}).  The radius of the
circumnuclear structure in NGC 1097 is about 550 pc (\cite{hum87}),
and about 450 pc in NGC 7469 (\cite{con91c}).  These are similar to
the one observed in NGC 3367 ($\sim 300$ pc).  None of these Seyfert
galaxies, however, show any indication of a large scale jet or
radio continuum extended lobes (\cite{hum87,wil91b}). In the case of normal barred galaxies,
like NGC 1326 and NGC 4314, they show H$\alpha$ circumnuclear structures
with radio continuum emission characteristic from regions of star formation
at similar radii (400 pc and 325 pc, respectively). None of them have radio
continuum emission from the compact nucleus (\cite{gar91a,gar91b}) nor any
radio continuum extended lobes. These differences in radio continuum emission
morphology from barred galaxies probably suggest also differences in
disk properties, like the amount of gas in the disk, the strength of the gravitational
potential including the non-axisymmetric component, and possibly different
ways of transporting material to the central regions and compact nucleus.

        Our main results are as follows:

1) There is a faint structure that connects the central
source with the SW lobe, here identified as a low surface brightness
jet.  Thus, the lobes are currently being fed by plasma from the compact radio
source.

2) At the highest resolution, the 8.4 GHz emission shows an
unresolved central peak with several circumnuclear sources.  The
unresolved source, which is located at the center of the galaxy, is
$\sim$ 10 times stronger than individual peaks of emission in the circumnuclear
region, and its deconvolved diameter is smaller than 65 pc.

3) The circumnuclear radio sources are most likely not associated with the interaction
of the jet and the sourrounding medium.

4) An estimate of the spectral index from the innermost region within  $3''.5$
is about -0.53.

5) The flow of plasma from the nuclear region to the lobes is out of the plane of
the galaxy but inclined with respect to the rotation axis of the disk.

6) No emission is detected from the stellar bar.

Our high resolution imaging has confirmed ideas from earlier work
that NGC3367 is a currently active low power radio galaxy, continuing
to be powered by a weak Liner/HII nucleus.
Like other barred spirals, it shows indications of star formation both
in the center and on larger scales. The jet-like structure
connecting the central source with the SW lobe is out of the plane of
the disk.

\section*{Acknowledgements}
It is a pleasure to thank the referee, Dr. Jim Ulvestad, for his
comments and suggestions on how to improve this paper.
We thank Drs. Barry Clark and W. Miller Goss for the
VLA allocated observing time.  JAG-B thanks the assistance and help of
Min Su Yun and Greg Taylor in the initial calibration of the VLA data,
and acknowledges partial financial support from DGAPA-UNAM and by
CONACYT (Mexico) that enabled him to work during  the period from 1997 to
1998 at the Department of Astronomy of the University of Minnesota where part
of the analysis was done.  JF acknowledges partial support from a
grant of DGAPA-UNAM. JAG-B and JF thank the hospitality of the 2001
Guillermo Haro Workshop, at INAOE, where part of this paper was
written.  LR extragalactic research at Minnesota is supported by the
National Science Foundation through grant NSF-AST 96-16984. We acknowledge
W. Wall for critical reading of the paper. This research
has made use of the NASA/IPAC Extragalactic Database (NED) which is
operated by the Jet Propulsion Laboratory, California Institute of
Technology under contract with the National Aeronautics and Space
Administration.

\normalsize

\clearpage
\newpage
\small

\begin{table}
\caption[ ]{
Radio Continuum Observations of NGC 3367}
\begin{flushleft}
\begin{tabular}{lcrcrc}
\hline
Frequency        &  Structure & Flux (mJy)  & log (P/W Hz$^{-1}$)   & Resolution   & References \cr
\hline
\cr
1490 MHz & Center    & 18.40$^a$ & 21.6 & $15''.0$ & 1 \cr
1490 MHz & SW lobe   & 14.70$^a$ & 21.5 & $15''.0$ & 1 \cr
1425 MHz & Center    & 15.80$^a$ & 21.5 & $4''.5$  & 2 \cr
1425 MHz & Triple    & 51.50$^b$ & 22.0 & $4''.5$  & 2 \cr
1425 MHz & Center    & 11.20$^a$ & 21.4 & $2''.1$  & 3 \cr
1425 MHz & Center    & 13.10$^c$ & 21.5 & $2''.1$  & 3 \cr
1425 MHz & Center    & 16.80$^d$ & 21.6 & $2''.1$  & 3 \cr
8460 MHz & Center    & 0.96$^a$  & 20.3 & $0''.3$  & 3 \cr
8460 MHz & Center    & 0.96$^e$  & 20.5 & $0''.3$  & 3 \cr
8460 MHz & Center    & 5.1$^c$   & 21.0 & $0''.3$  & 3  \cr
8460 MHz & Center    & 5.4$^f$   & 21.1 & $0''.3$  & 3 \cr

\hline

\end{tabular}
\end{flushleft}
$^a$ Peak flux.

$^b$ Peak flux from center plus integrated fluxes from lobes.

$^c$ Integrated flux from inner $3''.5$

$^d$ Integrated flux from inner $7''$.

$^4$ Integrated flux from inner $0''.5$.

$^f$ Integrated flux from inner $4''.5$

References: (1)\cite{con90}; (2) \cite{gar98}; (3) This work
\end{table}


\clearpage

\begin{thebibliography}{}


\bibitem[Baum et al.,\ 1993]{bau93}Baum, S.A., O'Dea, C.P., Dallacassa,
D, de Bruyn, A.G. \& Pedlar, A.1993, \apj, 419, 553
\bibitem[Beck et al. 1999]{bec99}Beck, R., Ehle, M., Shoutenkov, V.,
Shukurov, A. \& Sokoloff, D. 1999, Nature, 397, 324
\bibitem[Bennett et al. 1986]{ben86}Bennett, C.L., Lawrence, C.R.,
Burke, B.F., Hewitt, J.N. \& Mahoney, J.1986, \apjs, 61, 1
\bibitem[Caswell \& Wills 1967]{cas67}Caswell, J.L. \& Wills, D.
1967, \mnras, 135, 231
\bibitem[Colbert et al.,\ 1996]{col96}Colbert, E.J.M., Baum, S.A.,
Gallimore, J.F., O'Dea, C.P., Christensen, J.A. 1996, \apj, 467, 551
\bibitem[Condon 1987]{con87}Condon, J.J. 1987, \apjs, 65, 485
\bibitem[Condon et al. 1990]{con90}Condon, J.J., Helou, G., Sander,
D.B. \& Soifer, B.T. 1990, \apjs, 73, 359
\bibitem[Condon \& Broderick 1991]{con91a}Condon, J.J. \& Broderick,
J.J. 1991, \aj, 102, 1663
\bibitem[Condon, Frayer \& Broderick 1991]{con91b}Condon, J.J., Frayer, D.T.
\& Broderick, J.J. 1991, \aj, 101, 362
\bibitem[Condon et al. 1998]{con98}Condon, J.J., Cotton, W.D.,
Greisen, E.W., Yin, Q.F., Perley, R.A.,Taylor, G.B. \& Broderick, J.J. 1998, \aj, 115, 1693
\bibitem[Condon et al. 1991]{con91c}Condon, J.J., Huang, Z.-P., Yin,
Q.-F.\& Thuan, T.X. 1991, \apj, 378, 65
\bibitem[Condon 1992]{con92}Condon, J.J. 1992, \araa, 30, 575
\bibitem[Condon, Anderson \& Broderick 1995]{con95}Condon, J.J., Anderson, E. \& Broderick,
J.J. 1995, \aj, 109, 2318
\bibitem[de Bruyn \& Wilson 1978]{deb78}de Bruyn, A.G. \& Wilson,
A.S. 1978, \aap, 64, 433
\bibitem[Dunne et al. 2000]{dun00}Dunne, L., Eales, S., Edmunds, M.,
Ivison, R., Alexander, P. \& Clements, D.L.
2000, \mnras, 315, 115
\bibitem[Fabbiano, Kim \& Trinchieri 1992]{fab92}Fabbiano, G., Kim,
D.-W.\& Trinchieri, G. 1992, \apjs, 80, 531
\bibitem[Faranoff \& Riley 1974]{far74}Faranoff, B.L., \& Ryle, J.M.,
\mnras, 167, 31p
\bibitem[Garcia-Barreto et al. 1991a]{gar91a}Garcia-Barreto, J.A.,
Dettmar, R.-J., Combes, F., Gerin, M., \& Koribalski, B. 1991a,
Rev.Mexicana Astron. Astrofis., 22, 197
\bibitem[Garcia-Barreto et al. 1991b]{gar91b}Garcia-Barreto, J.A.,
Downes, D., Combes, F., Gerin, M., Magri, C.,
Carrasco, L.\& Cruz-Gonzalez, I. 1991b, \aap, 244, 257
\bibitem[Garcia-Barreto et al. 1993]{gar93}Garcia-Barreto, J.A.,
Carrillo, R., Klein, U. \& Dahlem, M. 1993, Rev.Mex.Astron.Astrofis. 25, 31
\bibitem[Garcia-Barreto, Franco \& Carrillo
1996]{gar96a}Garcia-Barreto, J.A., Franco, J. \& Carrillo, R. 1996,
\apj, 469, 138
\bibitem[Garcia-Barreto et al. 1996]{gar96b}Garcia-Barreto, J.A.,
Franco, J., Carrillo, R., Venegas, S. \& Escalante-Ramirez, B.
1996, Rev.Mexicana Astron. Astrofis., 32, 89
\bibitem[Garcia-Barreto et al. 1998]{gar98}Garcia-Barreto, J.A.,
Rudnick, L., Franco, J. \& Martos, M. 1998, \aj, 116, 111
\bibitem[Garcia-Barreto \& Rosado 2001]{gar01}Garcia-Barreto, J.A.
\&  Rosado, M.,  2001, \aj, 121, 2540
\bibitem[Genzel et al. 1995]{gen95}Genzel, R., Weitzel, L.,
Tacconi-Garman, L.E., Blietz, M., Cameron, M., Krabbe, A., Lutz, D.
\& Sternberg, A. 1995, \apj, 444, 129
\bibitem[Gerber \& Lamb 1994]{ger94} Gerber, R. A. \& Lamb, S. 1994, ApJ, 431, 604
\bibitem[Gioia et al. 1990]{gio90}Giogia, I.M., Maccacaro, T. Schild,
R.E., Wolter, A. Stocke, J.T., Morris,
S.L. \& Henry, J.P. 1990, \apjs, 72, 567
\bibitem[Gower, Scott \& Wills 1967]{gow67}Gower, J.F.R., Scott, P.F.
\& Wills, D. 1967, Mem. R.A.S., 71, 49
\bibitem[Grosb\o l 1985]{gro85}Grosb\o l, P.J. 1985, \aaps, 60, 261
\bibitem[Helou, Soifer \& Rowan-Robinson 1985]{hel85}Helou, G., Soifer, B.T. \&
Rowan-Robinson, M. 1985, \apj, 298, L7
\bibitem[Ho et al. 1997]{ho97}Ho, L.C., Filippenko, A.V., \&
Sargent, W. L. W. 1997, \apjs, 112, 315
\bibitem[Ho \& Ulvestad 2001]{ho01}Ho, L.C., \& Ulvestad , J.S. 2001, \apjs, 133, 77
\bibitem[Huchtmeier \& Seiradakis 1985]{huc85}Huchtmeier, W.K. \&
Seiradakis, J.H. 1985, \aap, 143, 216
\bibitem[Hummel 1981]{hum81}Hummel, E. 1981, \aap, 93, 93
\bibitem[Hummel, van der Hulst \& Keel 1987]{hum87} Hummel, E., van
der Hulst, J.M., \& Keel, W.C. 1987, \aap,
172, 32
\bibitem[Israel \& van der Hulst 1983]{isr83}Israel, F.P. \& van der
Hulst, J.M. 1983, \aj, 88, 1736
\bibitem[Kinney et al. 2000]{kin00}Kinney, A.L., Schmitt, H.R., Clarke, C.J.,
Pringle, J.E., Ulvestad, J.S. \& Antonucci, R.R.J. 2000, \apj, 537, 152
\bibitem[Lawrence et al. 1983]{law83}Lawrence, C.R., Bennett, C.L.
Garcia-Barreto, J.A., Greenfield, P.E. \& Burke, B.F. 1983, \apjs, 51, 67
\bibitem[Ledlow, Owen \& Keel 1998]{led98}Ledlow, M.J., Owen, F.N.,\&
Keel, W.C., 1998, \apj, 495, 227
\bibitem[Ledlow et al. 2001]{led01}Ledlow, M.J., Owen, F.N., Yun,
M.S., \& Hill, J.M., 2001, \apj, 552, 120
\bibitem[Lindblad 1999]{lin99}Lindblad, P.O. 1999, A\&ARv, 9, 221
\bibitem[Massaglia, Bodo \& Ferrari 1996]{mas96}Massaglia, S., Bodo,
G. \& Ferrari, A. 1996, \aap, 307, 997
\bibitem[Mauder et al. 1994]{mau94}Mauder, W., Weigelt, G.,
Appenzeller, I. \& Wagner, S.J. 1994, \aap, 285, 44
\bibitem[Miles, Houck \& Hayward 1994]{mil94}Miles, J.W., Houck, J.R.
\& Hayward, T.L. 1994, \apjl, 425, L37
\bibitem[Nagar \& Wilson 1999]{nag99}Nagar, N. M. \& Wilson, A.S.
1999, \apj, 516, 97
\bibitem[Niklas, Klein \& Wielebinski 1997]{nik97}Niklas, S. Klein, U.
\& Wielebinski, R. 1997, \aap, 322, 19
\bibitem[Ondrechen \& van der Hulst 1983]{ond83}Ondrechen, M.P. \&
van der Hulst, J.M. 1983, \apjl, 269, L47
\bibitem[Pringle et al. 1999]{pri99}Pringle, J.E., Antonucci, R.R.J., Clarke, C.J.,
Kinney, A.L., Schmitt, H.R. \& Ulvestad, J.S. 1999, \apj, 526, L9
\bibitem[Schilizzi et al. 2001]{sch01}Schilizzi, R.T., Tian, W.W.,
Conway, J.E., Nan, R., Miley, G.K., Barthel, P.D., Normandeau, M.,
Dallacasa, D. \& Gurvits, L.I. 2001, \aap, 368, 398
\bibitem[Schmitt et al. 1997]{sch97}Schmitt, H.R., Kinney, A.L.,
Storchi-Bergmann \& Antonucci, R. 1997, \apj, 477, 623
\bibitem[Schommer et al. 1988]{sch88}Schommer, R.A., Caldwell, N., Wilson, A.S.,
Baldwin, J.A., Phillips, M.M., Williams, T.B. \& Turtle, A.J. 1988,
\apj, 324, 154
\bibitem[Soifer et al. 1989]{soi89}Soifer, B.T., Bohemer, L., Neugebauer, G. \&
Sanders, D.B. 1989, \aj, 98, 766
\bibitem[Sramek 1975]{sra75}Sramek, R. 1975, \aj, 80, 771
\bibitem[Stocke et al. 1991]{sto91}Stocke, J.T., Morris, S.L.,
Giogia, I. M., Maccacaro, T. Schild, R., Wolter,
A. Fleming, T.A., \& Henry, J.P. 1991, \apjs, 76, 813
\bibitem[Tully 1988]{tul88}Tully,R.B. 1988, Nearby Galaxy Catalog,
(Cambridge:Cambridge Univ. Press)
\bibitem[Ulvestad \& Wilson 1984]{ulv84}Ulvestad, J. S. \& Wilson, 
A.S. 1984, \apj, 285, 439
\bibitem[Ulvestad, Neff \& Wilson 1987]{ulv87}Ulvestad, J.S., Neff, 
S.G. \& Wilson, A.S. 1987, \aj, 92, 22
\bibitem[Ulvestad \& Ho 2001]{ulv01}Ulvestad, J.S. \& Ho, L.C. 2001,
\apj, 558, 561
\bibitem[V\'eron-Cetty \& V\'eron 1986]{ver86}V\'eron-Cetty, M.-P.,
\& V\'eron, P. 1986, \aaps, 66, 335
\bibitem[V\'eron, Gon\c{c}alves \& V\'eron-Cetty 1997]{ver97}V\'eron, 
P., Gon\c{c}alves, A.C. \& V\'eron-Cetty,
M.-P., 1997, \aap, 319, 52
\bibitem[Wilson 1982]{wil82}Wilson, A.S. 1982 in Extragalactic Radio 
Sources, IAU 97, eds. D.S. Heeschen
\& C.M. Wade (Dordrecht: Kluwer Academic), p.179
\bibitem[Wilson 1991]{wil91a}Wilson, A.S. 1991 in The Interpretation 
of Modern Synthesis Observations of Spiral
Galaxies, ASP Conf.S. 18, eds. N. Duric \& P.C. Crane, p. 227
\bibitem[Wilson \& Colbert 1995]{wil95}Wilson, A.S., \& Colbert, E.
J. M., 1995, \apj, 438, 62
\bibitem[Wilson et al. 1991]{wil91b}Wilson, A. S., Helfer, T.T.,
Hanif, C. A. \& Ward, M.J. 1991, \apj, 381, 79
\bibitem[Wilson \& Ulvestad 1983]{wil83}Wilson, A.S., \& Ulvestad, J.
S., 1983, \apj, 275, 8
\bibitem[Wilson \& Ulvestad 1987]{wil87}Wilson, A.S., \& Ulvestad, J.
S., 1983, \apj, 319, 105

\end{thebibliography}
\end{document}